\documentclass[11pt]{article}
\usepackage{Blois,epsfig}
\def\be{\begin{equation}}
\def\ee{\end{equation}}
\def\bea{\begin{eqnarray}}
\def\eea{\end{eqnarray}}

\begin{document}
\vspace*{2cm}

\title{ Effects of Dynamical Quarks on the Stability of Heavy
Quarkonia in Quark-Gluon Plasma}

\author{Cheuk-Yin Wong}

\address{Physics Division, Oak Ridge National Laboratory, Oak Ridge, TN
37831\\
Department of Physics, University of Tennessee, Knoxville, TN 
37996}

\maketitle\abstracts{ To study the stability of heavy quarkonia in
quark-gluon plasma, we use a color-singlet $Q$-$\bar Q$ potential
determined previously to be a well-defined linear combination of the
free energy $F_1$ and the internal energy $U_1$.  Using the lattice
gauge results of Kaczmarek $et~al.$, the dissociation temperatures of
$J/\psi$ and $\chi_b$ in quenched QCD are found to be 1.62$T_c$ and
$1.18T_c$ respectively, in good agreement with spectral function
analyses.  The dissociation temperature of $J/\psi$ in full QCD with 2
flavors is 1.42$T_c$.  Thus, the presence of dynamical quarks in full
QCD lowers the dissociation temperature of $J/\psi$, but $J/\psi$
remains bound up to 1.42$T_c$.  }


\vspace*{-0.4cm}
\section{Introduction}
\vspace*{-0.2cm}

Previously, Matsui and Satz suggested that the suppression of $J/\psi$
production can be used as a diagnostic tool for quark-gluon plasma
\cite{Mat86}.  The stability of heavy quarkonia has been the subject
of many recent investigations $^{2-13}$.  Recent spectral analyses of
quarkonium correlators in lattice gauge calculations indicate that
$J/\psi$ is bound up to 1.6$T_c$ in quenched QCD \cite{Asa03,Dat03}.
Kaczmarek $et~al.$ have carried out lattice gauge calculations and
obtained the free energy $F_1$ and the total internal energy $U_1$ for
a color-singlet $Q$-$\bar Q$ pair in quark-gluon plasma, in both
quenched QCD and full QCD with 2 flavors \cite{Kac03,Kac05}.  It is of
interest to test whether the potential model using $F_1$ and $U_1$
lead to results consistent with spectral function analyses in the same
quenched approximation. If so, the potential model can be reliably
used in full QCD to assess the effects of dynamical quarks on the
stability of heavy quarkonia in quark-gluon plasma.

\vspace*{-0.2cm}
\section{ The $Q$-$\bar Q$ Potential and Heavy Quarkonium Stability}
\vspace*{-0.2cm}

The most important physical quantity in the potential model is the
$Q$-$\bar Q$ potential that acts between a quark $Q$ and an antiquark
$\bar Q$.  The potential should be obtained by proper theoretical
considerations and must yield results consistent with spectral
function analyses.

For a $Q$-$\bar Q$ pair in a thermalized system at a fixed temperature
and volume, the equation of motion for the $Q \bar Q$ pair can be
obtained by minimizing the grand potential with respect to the $Q \bar
Q$ wave function.  Using such a variational principle in a schematic
model, we find in Ref.\ [ \cite{Won04}] that the proper color-singlet
$Q$-$\bar Q$ potential is given by the internal energy $U_1$ after
subtracting out the quark-gluon plasma internal energy, as the
quark-gluon plasma internal energy gives rise to the potential for
the plasma constituents and not directly to the potential between
$Q$ and $\bar Q$.  Such a subtraction can be carried out by noting
that the quark-gluon plasma energy density $\epsilon$ is related to
its entropy density $\sigma$ and pressure $p$ by the first law of
thermodynamics, and the quark-gluon plasma energy density is also
related to the pressure by the equation of state.  Thus after taking
into account the response of the quark-gluon plasma to the polarizing
presence of the $Q$ and the $\bar Q$ in a hydrodynamical framework,
the proper $Q$-$\bar Q$ potential, $U_{Q\bar Q}^{(1)}$, is a linear
combination of the free energy $F_1$ and $U_1$ given by
\cite{Won04}
\begin{eqnarray}
\label{Uqq1}
U_{Q\bar Q}^{(1)}({\bf r},T)= 
 \frac{3}{3+a(T)}    F_1({\bf r},T) 
+\frac{a(T)}{3+a(T)} U_1({\bf r},T),
\end{eqnarray}
where $a(T)=3p(T)/\epsilon(T)$ can be obtained from the quark-gluon
plasma equation of state.  The potential $U_{Q\bar Q}^{(1)}$ is
approximately $F_1$ near $T_c$ and approximately $3F_1/4+U_1/4$ for $T
> 1.5 T_c$.

\vspace*{0.2cm}
\begin{figure} [h]
\includegraphics[angle=0,scale=0.43]{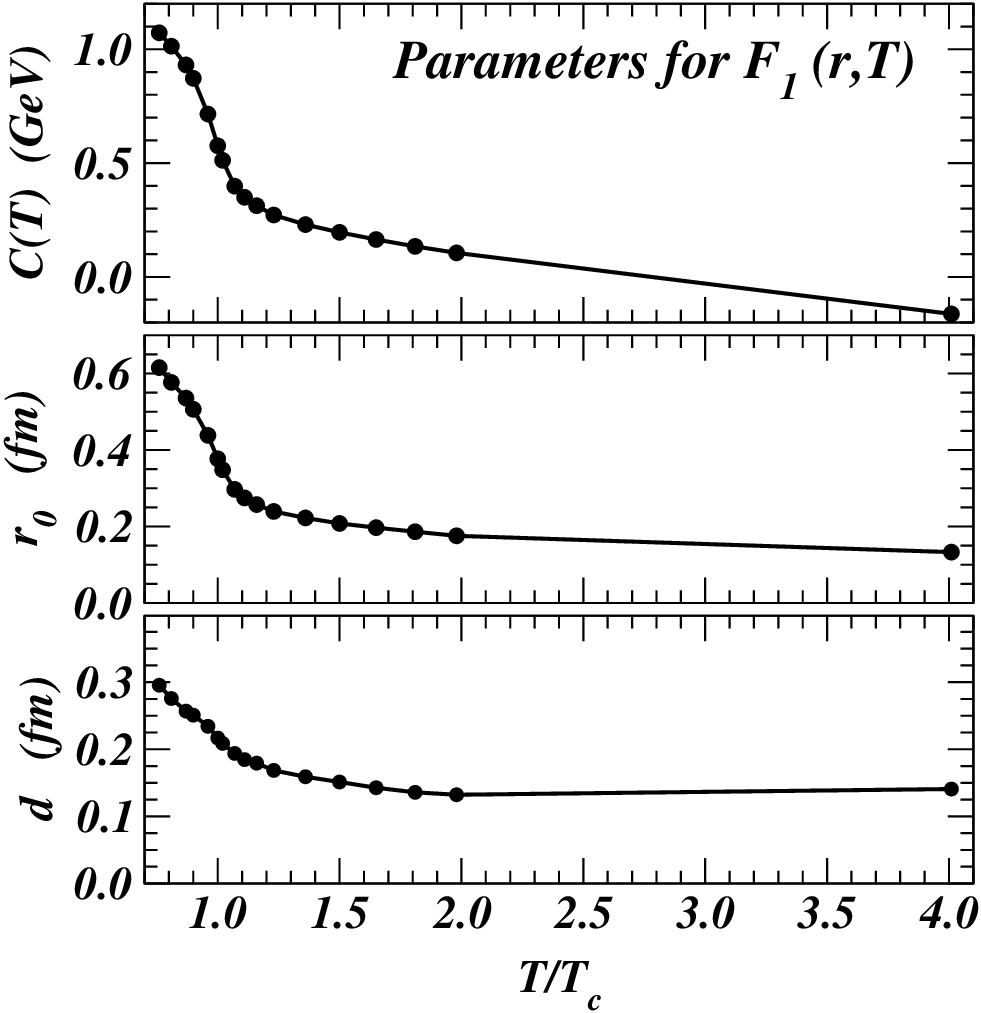}
\vspace*{1.3cm}\hspace*{-7.0cm}
\begin{minipage}{7.0cm}
\vspace*{0.3cm}\hspace*{-6.0cm}
\caption{Parameters for $F_1({\bf r},T)$ for full QCD.
}
\end{minipage}
\end{figure}

\vspace*{-10.9cm}\hspace*{10.0cm}
\begin{figure} [h]
\vspace*{0.0cm}\hspace*{7.7cm}
\begin{minipage}{8.0cm}
\includegraphics[angle=0,scale=0.47]{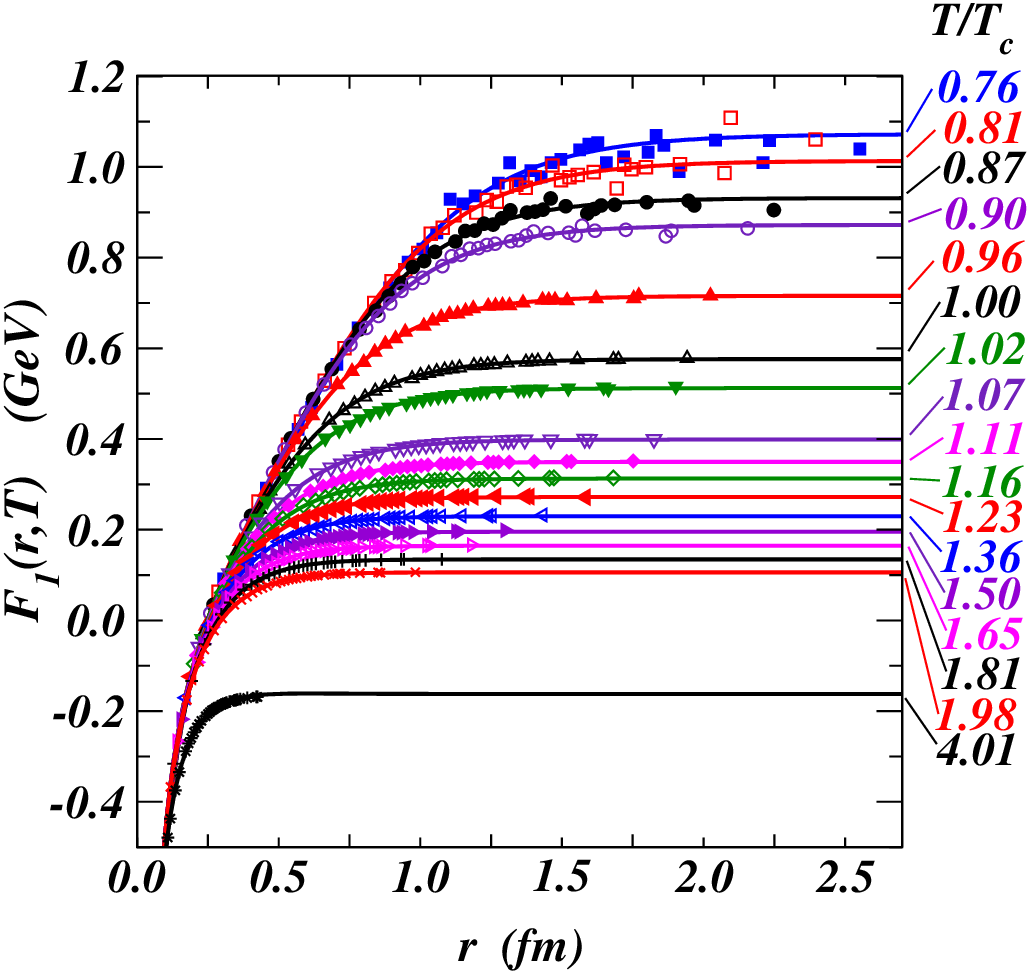}
\vspace*{-0.8cm}
\caption{Fits to $F_1({\bf r},T)$ of Ref.\ 
[ \protect\cite{Kac05}] for full QCD. 
}
\end{minipage}
\end{figure}

Kaczmarek $et~al.$ recently obtained $F_1$ and $U_1$ in both quenched
QCD \cite{Kac03} and full QCD with 2 flavors \cite{Kac05}.  In
quenched QCD, $F_1$ and $U_1$ can be parametrized in terms of a
screened Coulomb potential with parameters shown in Figs. 1 and 2 of
Ref. [~\cite{Won04}].  In full QCD with 2 flavors, $F_1$ and $U_1$ can
be simply represented by a color-Coulomb interaction at short
distances and a completely screened, constant, potential at large
distances, although other alternative representations have also been
presented \cite{Alb05,Dig05}.  The transition between the two
different spatial regions can be described by a radius parameter
$r_0(T)$ and a diffuseness parameter $d(T)$,
\begin{eqnarray}
\label{scoulF}
\{F_1,U_1\}({\bf r},T)=
-\frac{4}{3}\frac{\alpha_s(T)}{r}f(r,T)+C(T)[1-f(r,T)],
\end{eqnarray}
\vspace*{-0.6cm}
\begin{eqnarray}
\label{scoulU}
f(r,T)=\frac{1}{\exp \{(r-r_0(T))/d(T)\}+1 }.
\end{eqnarray}
In searching for the coupling constant $\alpha_s$, we found that the
value of $\alpha_s$ centers around 0.3.  It is convenient to fix
$\alpha_s$ to be 0.3 so that there are only three parameters for each
temperature.  The sets of parameters $C$, $r_0$, and $d$ for $F_1({\bf
r},T)$ and $U_1({\bf r},T)$ in full QCD with 2 flavors are shown in
Figs.\ 1 and 3 respectively, and the corresponding fits to the $F_1$
and $U_1$ lattice results of Kaczmarek $et~al.$ for full QCD with 2
flavors are shown in Figs. 2 and 4.  To calculate the ratio $a(T)$ in
Eq.\ (1), we use the quenched equation of state of Boyd
$et~al.$\cite{Boy96} for quenched QCD, and the equation of state of
Karsch $et~al.$\cite{Kar00} for full QCD with 2 flavors. Using quark
mases $m_c=1.4$ GeV and $m_b=4.3$ GeV, we can calculate the binding
energies of heavy quarkonia and their dissociation temperatures.  We
list in Table I the heavy quarkonium dissociation temperatures
calculated with the $U_{Q\bar Q}^{(1)}$ potential, the $F_1$
potential, and the $U_1$ potential, in both quenched QCD and full QCD.

\vspace*{0.0cm}
\begin{figure} [h]
\includegraphics[angle=0,scale=0.44]{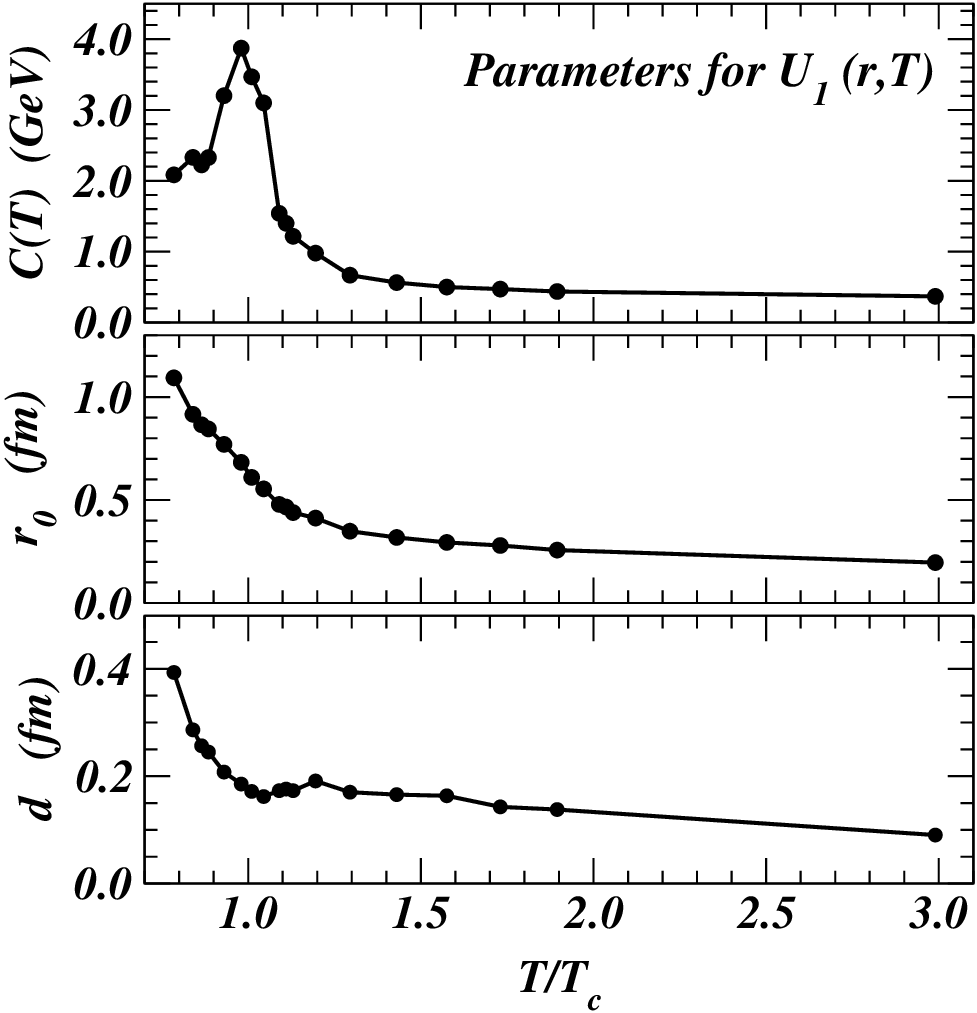}
\vspace*{1.5cm}\hspace*{-6.2cm}
\end{figure}
\vspace*{-11.0cm}\hspace*{10.0cm}
\begin{figure} [h]
\vspace*{0.0cm}\hspace*{8.5cm}
\begin{minipage}{8.0cm}
\includegraphics[angle=0,scale=0.41]{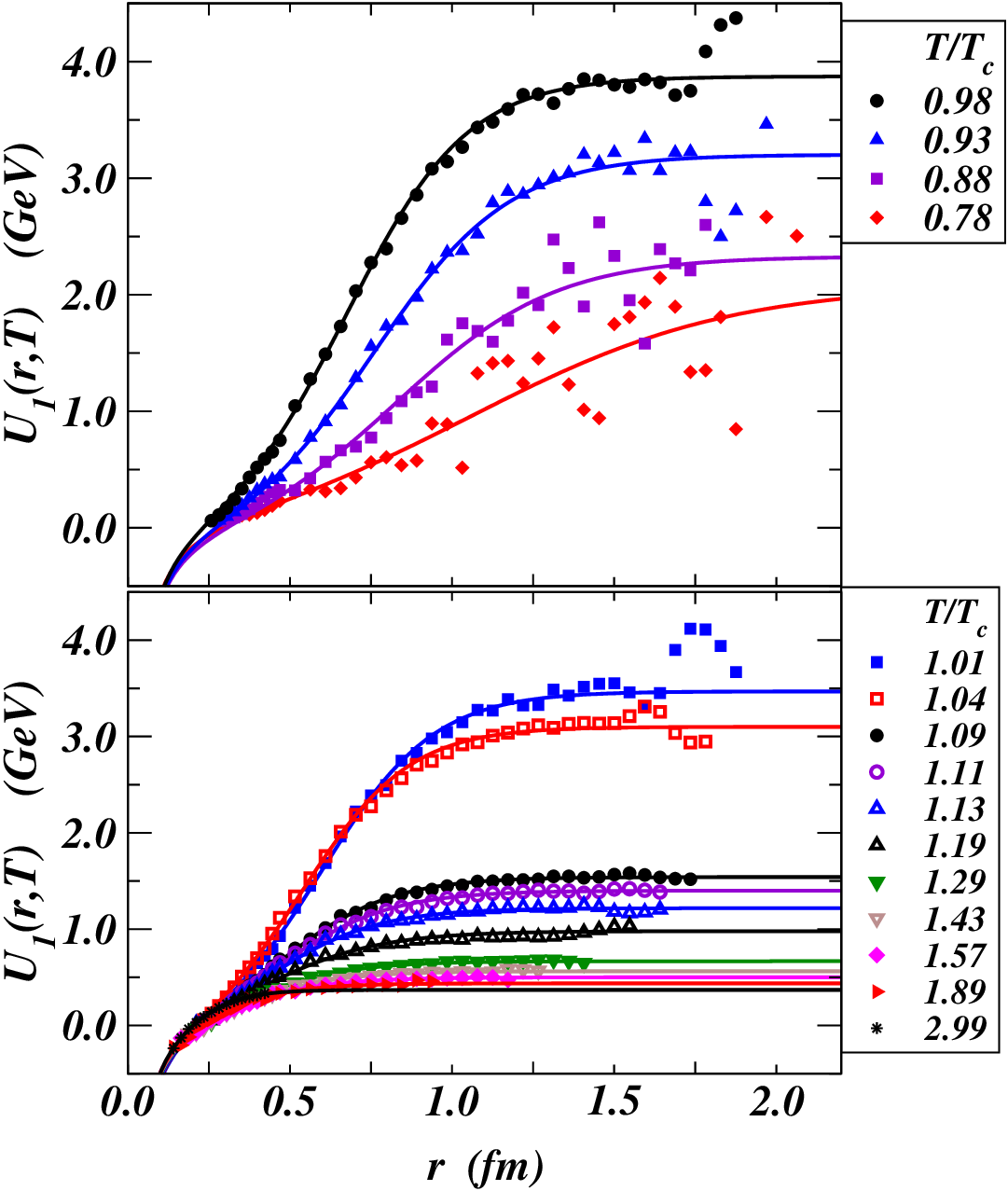}
\end{minipage}
\end{figure}

\vspace*{-0.5cm}
\noindent
{\small ~~~Figure 3. Parameters for $U_1({\bf r},T)$ for full QCD.~~~~~~~
Figure 4. Fits to $U_1({\bf r},T)$ of Ref.\ [ \cite{Kac05}] for full QCD.}

\vspace*{0.2cm} 
With the $U_{Q\bar Q}^{(1)}$ potential, the
dissociation temperatures of $J/\psi$ and $\chi_b$ are found to be
1.62$T_c$ and $1.18T_c$ respectively in quenched QCD.  On the other
hand, spectral analyses in quenched QCD give the dissociation
temperature of 1.62-1.70$T_c$ for $J/\psi$ \cite{Asa03} and
1.15-1.54$T_c$ for $\chi_b$ \cite{Pet05}.  Thus, the $U_{Q\bar
Q}^{(1)}$ potential of Eq.\ (\ref{Uqq1}) gives dissociation
temperatures that agree with those from spectral function
analyses.  In contrast, the $U_1$ potential in quenched QCD gives
dissociation temperatures that are much higher while the $F_1$
potential gives dissociation temperatures that are lower than the
corresponding temperatures from spectral function analyses.  Therefore
$U_{Q\bar Q}^{(1)}$ is the proper $Q$-$\bar Q$ potential for studying
the stability of heavy quarkonia in quark-gluon plasma.

\vskip 0.4cm \centerline{Table I.  Dissociation temperatures 
obtained from different analyses.} 
{\vskip 0.2cm\hskip -0.5cm
\begin{tabular}{|c|c|c|c|c|c|c|c|}
\hline
     & \multicolumn{4}{c|}{Quenched QCD} 
     & \multicolumn{3}{c|}{Full QCD (2 flavors)}       \\  \cline{2-7}
\hline
{\rm States     } & ~{\rm Spectral Anal.}~ 
                 &~ $U_{Q\bar Q}^{(1)}$
                 &  ~ $F_1$  & ~ $U_1$ 
                 &~ $U_{Q\bar Q}^{(1)}$ 
                 &  ~ $F_1$  & ~ $U_1$ \\

\hline
$J/\psi,\eta_c$       &   $1.62$-$1.70T_c^\dagger$ 
&   $1.62\,T_c$  
&   $1.40 \,T_c$ & $2.60 \,T_c$              
&   $1.42\,T_c$  
&   $1.21 \,T_c$ & $2.22 \,T_c$              
\\ \cline{1-8}  
$\chi_c$              & below $ 1.1T_c^{\natural}$   
&   unbound
& unbound             &  $1.18 \,T_c$            
&   $1.05 \,T_c$
&   unbound           &  $1.17 \,T_c$            
\\ \cline{1-8}
$\psi',\eta_c'$       &  
&   unbound 
&   unbound & $1.23 \,T_c$              
&   unbound  
&   unbound & $1.11 \,T_c$              
\\ \cline{1-8}  
$\Upsilon,\eta_b$     &                   
&  $ 4.1 \,T_c$    
&    $ 3.5 \,T_c$   &    $ \sim 5.0  \,T_c$                      
&  $ 3.30 \,T_c$    
&    $ 2.90 \,T_c$   &    $ 4.18 \,T_c$                           
\\ \cline{1-8}  
$\chi_b$              &  $1.15$-$1.54T_c^{\sharp}$     
&  $ 1.18 \,T_c$      
&  $ 1.10 \,T_c$      & $ 1.73 \,T_c$            
&  $ 1.22 \,T_c$      
&    $ 1.07 \,T_c$    & $ 1.61 \,T_c$            
\\ \cline{1-8}
$\Upsilon',\eta_b'$     &                   
&  $ 1.38 \,T_c$    
&    $ 1.19  \,T_c$   &    $ 2.28  \,T_c$                      
&  $ 1.18 \,T_c$    
&    $ 1.06 \,T_c$   &    $ 1.47 \,T_c$                           
\\ \cline{1-8}
                                                        \hline
\multicolumn{8}{l}
{${}^\dagger$Ref.\cite{Asa03},~~~     
 ${}^{\natural}$Ref.\cite{Dat03},~~~     
 ${}^{\sharp }$Ref.\cite{Pet05}}     
\end{tabular}
}

\vspace*{0.2cm}
In full QCD with 2 flavors, the dissociation temperature of $J/\psi$
calculated with the $U_{Q\bar Q}^{(1)}$ potential is 1.42$T_c$, as
listed in Table I.  Thus, the presence of dynamical quarks in full QCD
lowers the dissociation temperature of $J/\psi$ because of the
additional screening, but $J/\psi$ remains bound up to 1.42$T_c$.

\vspace*{-0.2cm}
\section{Discussions and Conclusions}
\vspace*{-0.2cm}

The stability of quarkonium depends on the potential that acts between
the quark $Q$ and an antiquark $\bar Q$.  Previous work in the
potential model uses the free energy $F_1$ \cite{Dig01a,Won01a,Bla05}
or the internal energy $U_1$ \cite{Kac02,Alb05} obtained from lattice
gauge calculations as the $Q$-$\bar Q$ potential.

We find in Ref.\ [ \cite{Won04}] that the proper $Q$-$\bar Q$
potential, $U_{Q\bar Q}^{(1)}$, is actually a well-defined linear
combination of $F_1$ and $U_1$ with coefficients that depend on the
equation of state.  We test such a $Q$-$\bar Q$ potential
by evaluating the dissociation temperatures for heavy quarkonia.
Within the same quenched approximation, the $U_{Q\bar Q}^{(1)}$
potential give results that are consistent with those from spectral
function analyses. We can therefore use the $U_{Q\bar Q}^{(1)}$
potential in full QCD to study the effects of dynamical quarks on the
stability of heavy quarkonia.  

We find that for $J/\psi$, the dissociation temperature is changed
from 1.62$T_c$ for quenched QCD to 1.42$T_c$ for full QCD with 2
flavors.  Thus, the presence of dynamical quarks in full QCD lowers
the dissociation temperature of $J/\psi$ because of the additional
screening due to dynamical quarks.  The change in stability is
nonetheless quite small, as $J/\psi$ remains bound up to 1.42$T_c$
even in the presence of dynamical quarks.

In conclusion, we have shown that the potential model is consistent
with the spectral function analysis in quenched QCD, and we have found
that the effects of dynamical quarks lowers the $J/\psi$ dissociation
temperature to 1.42$T_c$.

\vspace*{-0.2cm}
\section*{Acknowledgments}
\vspace*{-0.1cm}

The author would like to thank Dr. O. Kaczmarek for sending him tables
of $F_1$ and $U_1$ from the lattice gauge calculations.  The author
thanks Dr. H. Crater for helpful discussions.  This research was
supported in part by the Division of Nuclear Physics, U.S. Department
of Energy, under Contract No. DE-AC05-00OR22725, managed by
UT-Battelle, LLC and by the National Science Foundation under contract
NSF-Phy-0244786 at the University of Tennessee.

\end{document}